# Implication of family non-universal $Z'$ model to rare exclusive $b \to s(l\bar{l}, \nu\bar{\nu})$ transitions

P. Maji[1], P. Nayek[2] and S. Sahoo[3]
National Institute of Technology, Durgapur-713209, West Bengal, India
[1]Email: majipriya@gmail.com, [3]Email: sukadevsahoo@yahoo.com

**Abstract:** We have investigated $B \to (K, K^*)(l\bar{l}, \nu\bar{\nu})$ decay channels and predicted some form factor dependent observables (e.g. branching ratio, forward-backward asymmetry and lepton polarization asymmetry) considering new physics contribution from non-universal $Z'$ model. We have found that there is appreciable difference between the standard model and $Z'$ model predictions. Recently, the lepton flavour non-universality parameters $R_K$ and $R_{K^*}$ have been measured by the LHCb collaboration and are found to be different from their standard model predictions. These deviations provide the hint of flavor non-universality in lepton sector and motivate our research to look for new physics beyond the standard model. The $R_K$ and $R_{K^*}$ anomalies may be due to new physics in either electron or muon sector or both. In our work, we have shown that the experimental results of $R_K$ and $R_{K^*}$ could be achieved in non-universal $Z'$ model considering the new physics couplings to muon sector only.

# Contents:

## 1    Introduction

With the discovery of Higgs boson, the standard model (SM) could be considered as the most successful theory for particle physics. It is well established experimentally. But due to some basic shortcomings like absence of gravity, hierarchy problem of CKM matrix, existence of massive neutrinos etc. the SM could not be regarded as the fundamental theory yet. To overcome these deficiencies one needs to go beyond SM in search of new physics (NP). In this context, rare B meson decays are holding up the dynamic role in recent times. The rare decay modes of B meson provide a stringent way to test the SM [1] and are also able to explore small contribution from new virtual particles. These channels are forbidden at tree level in the SM but allowed through GIM mechanism [2] at loop level. To probe NP, the decays which involve $b \to s$ transitions [3, 4] provide a number of observables such as forward-backward asymmetry, helicity fractions, single and double lepton polarization etc [5]. In literature, various inclusive [6] and exclusive [7] semiletonic B meson decay modes like $B \to X_{s,d} l^+ l^-$ and $B \to M l^+ l^-$ (where $M = K, K^*, K_1, \rho$ etc.) have been studied in a generous way. These rare semileptonic B decays induced by quark level flavour-changing neutral current (FCNC) $b \to s$ transitions are relatively clean compare to the pure hadronic transitions and also expected to be sensitive to possible new interactions, e.g. supersymmetric theories [8], 2HDM [9], top-color [10], leptoquarks [11,12], non-universal $Z'$ boson [13,14] etc. The response to various NP models could be utilized to explain the anomalies which are coming to the picture from recent experiments.

Recently, among all rare decays the semileptonic channel $B \to K^*(\to K\pi)\mu\bar{\mu}$ has gained the main interest due to a huge variety of the discrepancies in its angular observables. Various B physics experiments such as BaBar [15], Belle [16], CDF [17] and LHCb [18] have provided data for the measurements of angular distributions of $B \to (K, K^*) l^+ l^-$ decay processes both in low and high recoil regime except the intermediate region for $q^2 \sim m_{J/\psi}^2$, where, $q^2$ is the invariant mass squared of the lepton pair. The middle region is dominated by charmonium resonance background introducing large hadronic uncertainties to many of the parameters. One can use QCD factorization for measuring different physical observables in high recoil of momentum. Whereas simultaneous involvement of heavy quark effective theory and operator product expansion in $1/Q$ with $Q = (m_b, q^2)$ could lead to a reliable computation of angular observables in low recoil. From the experimental point of view, LHCb found that there is a sizable deviation from SM prediction in $P_5'$ observable which is in conflict of almost 3.7σ with 1fb$^{-1}$ luminosity at large recoil $s \in [4.30, 8.68]$ [19], where $s$ is the dimuon invariant mass squared $q^2$. Another noticeable anomaly found by LHCb lies in the current measurement of the ratios $R_K = BR(B \to K\mu^+\mu^-)/BR(B \to Ke^+e^-)$ and $R_{K^*} = BR(B \to K^*\mu^+\mu^-)/BR(B \to K^*e^+e^-)$. Within SM, these ratios are unity. But the recent experiments [20, 21] have observed that

$$R_K = 0.745^{+0.090}_{-0.074}(\text{stat}) \pm 0.036(\text{sys}), \text{ for } 1.1 < q^2 < 6 \text{ GeV}^2.$$

$$R_{K^*} = \begin{cases} 0.66^{+0.11}_{-0.07}(\text{stat}) \pm 0.03(\text{sys}), \text{ for } 0.045 < q^2 < 1.1 \text{ GeV}^2, \\ 0.69^{+0.11}_{-0.07}(\text{stat}) \pm 0.05(\text{sys}), \text{ for } 1.1 < q^2 < 6 \text{ GeV}^2. \end{cases}$$

Thus, there is 2.6σ deviation of $R_K$ within the range $1 < q^2 < 6$ GeV$^2$/c$^4$ and for $R_{K^*}$, the discrepancy is 2.1σ -2.3σ for low-$q^2$ region and 2.4σ-2.5σ for central-$q^2$ region from the SM prediction. When the measurements of $R_K$ and $R_{K^*}$ are combined, a deviation around 3σ is found. and that provides strong hint of lepton flavour universality violation [22]. From theoretical point of view $R_K$ and $R_{K^*}$ are very clean observables as the hadronic uncertainties are cancelled out in the ratios. So, the experimental results are restricting the bound of NP. Another discrepancy observed by both BaBar [15] and Belle [16] collaborations that their fitted forward-backward asymmetry (A$_{FB}$) spectrum is generally higher than the SM expectation in all $q^2$ region. The experimental measurement shows

$$A_{FB}(B \to K^* l^+ l^-)_{0\text{GeV}^2 \le q^2 \le 2\text{GeV}^2} = 0.47^{+0.26}_{-0.32} \pm 0.03.$$

The rare semileptonic decay modes of B meson having $\nu\bar{\nu}$ as final state particles, such as, $B \to (K, K^*)\nu\bar{\nu}$ [23] are also relevantly suppressed within the SM and their long-distance contributions are not much prior generally. These decays are very clean theoretically due to the absence of of photonic penguin contributions and strong suppression of light quarks. In current date, the inclusive channels of these decay modes are probably out of reach due to missing neutrinos but the exclusive decays like $B \to (K, K^*)\nu\bar{\nu}$ are more promising for measuring branching ratios and other related observables. Theoretical study of these decays requires significant form factors calculated from non-perturbative approach.

Experimentally observed discrepancies for the exculsive decay channels of B meson have motivated many researchers to investigate whether the deviations are coming from unknown factorisable power corrections or from NP. After taking into account both the arguments, global analyses [4, 24, 25] are putting a strong benchmark for the presence of NP to explain these anomalies. One convenient way to analyze the effects of NP in the observables and also to calibrate them with SM predictions is through the effective Hamiltonian approach which allows model-independent study of NP [26-28]. Another possibility is to compute this effective Hamiltonian with a background of particular NP model [29, 30].

At tree level there are very few candidates which can be utilized in probing the NP responsible for B anomalies. These are basically leptoquark [31-33], which could be either scalar or vector, and $Z'$ vector boson [34, 35]. Both of them are quite significant in explaining various NP phenomena. In particular to our study, we have picked up the non-universal $Z'$ model [36-41] which is considered to be the most economical and alleviate most of the existing anomalies simultaneously. This model is economical in the sense that it requires one extra $U(1)'$ gauge symmetry associated with a neutral gauge boson called $Z'$ boson. The main attraction of this model is drawn to the fact that FCNC transitions could occur at tree level due to the off-diagonal couplings of non-universal $Z'$ with fermions. Various studies of non-universal $Z'$ model have been done and it is observed that it can help to resolve the puzzles of rare B meson decays such as $B - \bar{B}$ mixing phase [42, 43], $\pi - K$ puzzle [44] etc.

In this paper, we intend to study the effect of non-universal $Z'$ model on different observables of $B \to (K, K^*)l\bar{l}$ and $B \to (K, K^*)\nu\bar{\nu}$ decay modes. To develop an accurate platform for the form factors for whole kinematical accessible region the author of ref. [45] has approached the dispersion quark model. This model provides the correct heavy-quark expansion at leading and next-to-leading $1/m_Q$ orders in accordance with QCD for transitions between heavy quark [46, 47]. For heavy to light quark transitions the form factors provide some fascinating relations with vector, axial-vector and tensor currents which are valid in low recoil [48]. In this way the form factors of the dispersion quark model obey all severe constraints coming from theoretical restrictions. Formulas derived from this quark model satisfy the SM predicted values of differential decay rates and asymmetries for $B \to (K, K^*)(l^+l^-, \nu\bar{\nu})$ decays. Measurement of forward-backward asymmetry not only detects the NP beyond SM but also indicates the Lorentz structure present in it. Observed distribution of lepton polarization asymmetry and the shape of forward-backward asymmetry are mostly independent of the long-distance contributions. From such aspects both forward-backward asymmetry ($A_{FB}$) and lepton polarization asymmetry ($P_L$) are favouring a powerful way to test SM as well as to probe possible NP.

The paper is organized as follows. In section 2, we present the effective Hamiltonian responsible for $b \to sl^+l^-$ and $b \to s\nu\bar{\nu}$ processes. We also discuss the new physics contributions to the Hamiltonian due to the coupling of $Z'$ with quarks and leptons. In section 3, we discuss the matrix elements and form factors relevant to $B \to (K, K^*)$ decays and also presented the explicit expressions for different physical observables. In section 4, we have computed the branching ratios, forward-backward asymmetry and lepton polarization asymmetry with taking into account the contribution of $Z'$. We have also discussed the anomalies of lepton non-universality factors $R_K$ and $R_{K^*}$. Section 5 contains summary and conclusion.

## 2  Theoretical Framework

### i)  $b \to s(l\bar{l}, \nu\bar{\nu})$ decays in the Standard Model (SM)

In the SM neglecting the doubly Cabibo-supressed contributions, the effective Hamiltonian describing the $b \to sl^+l^-$ transitions can be written as [49]

$$\mathcal{H}_{eff} = \frac{G_F}{\sqrt{2}} V_{tb} V_{ts}^* \sum_i C_i(\mu) O_i(\mu) \,, \qquad (1)$$

where $G_F$ is the universal Fermi constant, $C_i$'s are Wilson coefficients and $O_i$'s are the basis operators [27]. In particular to the rare decays of our interest, the operators providing the main contribution within the SM are,

$$O_1 = (\bar{s}_\alpha \gamma^\mu (1-\gamma_5) b_\alpha)(\bar{c}_\beta \gamma_\mu (1-\gamma_5) c_\beta),$$
$$O_2 = (\bar{s}_\alpha \gamma^\mu (1-\gamma_5) b_\beta)(\bar{c}_\beta \gamma_\mu (1-\gamma_5) c_\alpha),$$

$$O_{7\gamma} = \frac{e}{8\pi^2} \bar{s}_\alpha \sigma_{\mu\nu} [m_b(\mu)(1+\gamma_5) + m_s(\mu)(1-\gamma_5)] b_\alpha F^{\mu\nu},$$
$$O_{9V} = \frac{e}{8\pi^2} (\bar{s}_\alpha \gamma^\mu (1-\gamma_5) b_\alpha) \bar{l} \gamma_\mu l,$$
$$O_{10A} = \frac{e}{8\pi^2} (\bar{s}_\alpha \gamma^\mu (1-\gamma_5) b_\alpha) \bar{l} \gamma_\mu \gamma_5 l. \qquad (2)$$

Here, $\mu$ is the renormalization scale which is usually fixed at $\mu \simeq m_b$ in order to avoid large logarithms in matrix elements of the operators $O_i$. The Wilson coefficients are calculated at large scale for $\mu \simeq M_W$ and then renormalized to the scale $\mu \simeq m_b$. The expressions for $C_i(\mu)$'s in the SM are given in ref. [50]. For our calculation we have taken $\mu = 4.8$ GeV and corresponding values of Wilson coefficients are as follows [51]:

**Table 1:** Values of Wilson coefficients

| $C_1$ | $C_2$ | $C_3$ | $C_4$ | $C_5$ | $C_6$ | $C_7^{eff}$ | $C_8^{eff}$ | $C_9$ | $C_{10}$ |
|---|---|---|---|---|---|---|---|---|---|
| -0.257 | 1.009 | -0.005 | -0.078 | 0.000 | 0.001 | -0.304 | -0.167 | 4.211 | -4.103 |

The four quark operators $O_1$ and $O_2$ provide both short- and long-distance contributions to the effective Hamiltonian. These contributions can be quantified if we replace $C_{9V}(m_b)$ by an effective coefficient $C_{9V}^{eff}(m_b, q^2)$ given by

$$C_{9V}^{eff}(m_b, q^2) = C_{9V}(m_b) + Y(m_b, q^2)^{pert} + Y(m_b, q^2)^{res}. \qquad (3)$$

The short-distance contributions are mainly lies in the perturbative part which is due to one loop correction in the matrix elements of four-quark operators. Besides that, $Y(m_b, q^2)^{pert}$ also receives long-distance contributions which arise due to the resonance term of intermediate $c\bar{c}$ loop, i.e. $Y(m_b, q^2)^{res}$. The analytic expressions for $Y(m_b, q^2)^{pert}$ and $Y(m_b, q^2)^{res}$ could be found in ref. [51] and [8] respectively.

After replacing the coefficient successfully, the above Hamiltonian leads to the following free quark matrix element [10, 52]

$$\mathcal{H}_{eff}(b \to sl^+l^-)$$
$$= \frac{G_F}{\sqrt{2}} \frac{\alpha_{em}}{2\pi} V_{tb} V_{ts}^* \left[ -2 \frac{C_{7\gamma}(m_b)}{q^2} \left( (m_b + m_s)(\bar{s} i \sigma_{\mu\nu} q^\nu b) \right. \right.$$
$$\left. + (m_b - m_s)(\bar{s} i \sigma_{\mu\nu} q^\nu \gamma_5 b) \right) (\bar{l} \gamma^\mu l) + C_{9V}^{eff}(m_b, q^2)(\bar{s}\gamma_\mu(1-\gamma_5)b)(\bar{l}\gamma^\mu l)$$
$$\left. + C_{10A}(m_b)(\bar{s}\gamma_\mu(1-\gamma_5)b)(\bar{l}\gamma^\mu \gamma_5 l) \right], \qquad (4)$$

$$\mathcal{H}_{eff}(b \to s\nu\bar{\nu}) = \frac{G_F}{\sqrt{2}} \frac{\alpha_{em}}{2\pi \sin^2\theta_W} V_{tb} V_{ts}^* X(x_t) (\bar{s}_\alpha \gamma_\mu (1-\gamma_5) b^\alpha)(\bar{\nu}\gamma^\mu(1-\gamma_5)\nu), \qquad (5)$$

where, $x_t = (m_t/M_W)^2$ and $\theta_W$ is Weinberg's angle. The expression for $X(x_t)$ is given in ref. [52, 53].

### ii) $b \to s(l\bar{l}, \nu\bar{\nu})$ decays in family non-universal $Z'$ model

Several studies have been done considering universal $Z'$ couplings with the three families of fermions [54, 55]. The reason behind this is both theoretical and phenomenological. Considering the universal couplings, one can easily avoid the FCNC mediated transitions at tree level $Z'$ exchange. But recent experimental results for decay rates of semileptonic B decays violate $\mu - e$ universality and also deviate from the SM prediction. This motivates us to consider the non-universal couplings of $Z'$ [38] with each of the fermion family.

In the presence of non-universal $Z'$, Hamiltonian for FCNC transitions could be written as following [38, 42, 56]

$$\mathcal{H}_{eff}^{Z'} = -\frac{4G_F}{\sqrt{2}} V_{tb} V_{ts}^* [\Lambda_{sb} C_9^{Z'} O_9 + \Lambda_{sb} C_{10}^{Z'} O_{10}], \tag{6}$$

where, $\Lambda_{sb} = \frac{4\pi e^{-i\phi_{sb}}}{\alpha_{em} V_{tb} V_{ts}^*}$, $C_9^{Z'} = |B_{sb}| S_{ll}^{LR}$, $C_{10}^{Z'} = |B_{sb}| D_{ll}^{LR}$

with, $S_{ll}^{LR} = B_{ll}^L + B_{ll}^R$, $D_{ll}^{LR} = B_{ll}^L - B_{ll}^R$.

Here, $B_{sb}$ corresponds to off diagonal left-handed coupling of $Z'$ with quarks, $B_{ll}^L$ and $B_{ll}^R$ are left- and right-handed couplings for $Z'$ with leptons. $\phi_{sb}$ is the new weak phase angle. These couplings are strongly constrained from low-energy experiments. The most useful feature of $Z'$ model is that the operator basis remains same as in the SM; only the modifications are done for $C_9$ and $C_{10}$ while $C_7^{eff}$ remains unchanged. The new Wilson coefficients $C_9$ and $C_{10}$ with the total contributions of SM and $Z'$ model are written as

$$C_9^{Total} = C_9^{eff} + C_9^{NP}$$

$$C_{10}^{Total} = C_{10} + C_{10}^{NP} \tag{7}$$

where, $C_9^{NP} = \Lambda_{sb} C_9^{Z'}$ and $C_{10}^{NP} = \Lambda_{sb} C_{10}^{Z'}$.

The numerical values of the $Z'$ couplings suffer from several constraints which arise due to different exclusive and inclusive B decays have been found in the ref. [43, 57]. We have used three scenarios in our calculation, corresponding to three different fitting values of $B_s - \bar{B}_s$ mixing data which present the couplings as well as the weak phase angle. The values of input parameters are set by UTfit collaborations [58] and recollected in Table 2.

The right handed couplings of $Z'$ with neutrinos are considered to be zero. So only the left handed couplings are present for $B \to (K, K^*)\nu\bar{\nu}$ decay modes. We have assumed the couplings of neutrino to $Z'$ are similar as the couplings of leptons and $Z'$. For the measurement of branching ratios of $B \to (K, K^*)\nu\bar{\nu}$ decays within $Z'$ model, the couplings are as follows:

$$B_{\nu\bar{\nu}}^R = 0, B_{ll}^L = B_{\nu\bar{\nu}}^L \quad \text{and} \quad S_{ll}^{LR} = D_{ll}^{LR} = S_{\nu\bar{\nu}}^{LR} = D_{\nu\bar{\nu}}^{LR}. \tag{8}$$

It is clear from the above relations that, $C_9^{Z'}$ and $C_{10}^{Z'}$ are same for the decays which include neutrinos. The couplings are universal for all three generations of neutrino. So, the determination of the neutrino type is beyond the discussion of this paper.

**Table 2:** The numerical values of the $Z'$ parameters [59]

|  | $|B_{sb}| \times 10^{-3}$ | $\phi_{sb}$(Degree) | $S_{LL} \times 10^{-2}$ | $D_{LL} \times 10^{-2}$ |
|---|---|---|---|---|
| $S_1$ | 1.09 ± 0.22 | -72 ± 7 | -2.8 ± 3.9 | -6.7 ± 2.6 |
| $S_2$ | 2.20 ± 0.15 | -82 ± 4 | 1.2 ± 1.4 | -2.5 ± 0.9 |
| $S_3$ | 4.0 ± 1.5 | 150 ± 10 (or -150 ± 10) | 0.8 | -2.6 |

## 3     Formulation for the Analysis
### i)     Matrix elements and form factors for $B \to (K, K^*)$ decays

The long-distance contribution to the $B \to (K, K^*)$ decays is associated with the meson matrix elements of bilinear quark currents of the SM Hamiltonian, which contains relativistic invariant transition form factors of vector, axial-vector and tensor currents. Weak annihilation is another mode to contribute the long-distance effect in rare semileptonic decays. This process is caused by Cabibo-suppressed part of four fermion operators. But, in $B \to (K, K^*)$ decays, the impact of this process is very small, so we have neglected it in our effective Hamiltonian. The currents mainly responsible for meson decays induced by $b \to s$ quark transitions are: vector current $V_\mu = \bar{s}\gamma_\mu b$, axial-vector current $A_\mu = \bar{s}\gamma_\mu \gamma^5 b$, tensor current $T_{\mu\nu} = \bar{q}\sigma_{\mu\nu} b$ and pseudo-tensor current $T_{\mu\nu}^5 = \bar{q}\sigma_{\mu\nu}\gamma_5 b$. The amplitudes for meson decays corresponding to these currents are given as

$$\begin{aligned}
\langle P(M_2, p_2)|V_\mu(0)|P(M_1, p_1)\rangle &= f_+(q^2)P_\mu + f_-(q^2)q_\mu, \\
\langle V(M_2, p_2, \epsilon)|V_\mu(0)|P(M_1, p_1)\rangle &= 2g(q^2)\epsilon_{\mu\nu\alpha\beta}\epsilon^{*\nu}p_1^\alpha p_2^\beta, \\
\langle V(M_2, p_2, \epsilon)|A_\mu(0)|P(M_1, p_1)\rangle &= i\epsilon^{*\alpha}[f(q^2)g_{\mu\alpha} + a_+(q^2)p_{1\alpha}P_\mu + a_-(q^2)p_{1\alpha}q_\mu], \\
\langle P(M_2, p_2)|T_{\mu\nu}(0)|P(M_1, p_1)\rangle &= -2is(q^2)(p_{1\mu}p_{2\nu} - p_{1\nu}p_{2\mu}), \\
\langle V(M_2, p_2, \epsilon)|T_{\mu\nu}(0)|P(M_1, p_1)\rangle &\\
&= i\epsilon^{*\alpha}[g_+(q^2)\epsilon_{\mu\nu\alpha\beta}P^\beta + g_-(q^2)\epsilon_{\mu\nu\alpha\beta}q^\beta + g_0(q^2)p_{1\alpha}\epsilon_{\mu\nu\beta\gamma}p_1^\beta p_2^\gamma], \\
\langle P(M_2, p_2)|T_{\mu\nu}^5(0)|P(M_1, p_1)\rangle &= s(q^2)\epsilon_{\mu\nu\alpha\beta}P^\alpha q^\beta, \\
\langle V(M_2, p_2, \epsilon)|T_{\mu\nu}^5(0)|P(M_1, p_1)\rangle &= g_+(q^2)(\epsilon_\nu^* P_\mu - \epsilon_\mu^* P_\nu) + g_-(q^2)(\epsilon_\nu^* q_\mu - \epsilon_\mu^* q_\nu) + \\
g_0(q^2)(p_1\epsilon^*)(p_{1\nu}p_{2\mu} - p_{1\mu}p_{2\nu}), & \quad (9)
\end{aligned}$$

where, $q = p_1 - p_2$, $P = p_1 - p_2$. We have used the following notations: $\gamma^5 = i\gamma^0\gamma^1\gamma^2\gamma^3$, $\sigma_{\mu\nu} = \frac{i}{2}[\gamma_\mu, \gamma_\nu]$, $\epsilon^{0123} = -1$, $\gamma_5\sigma_{\mu\nu} = \frac{i}{2}\epsilon_{\mu\nu\alpha\beta}\sigma^{\alpha\beta}$ and $Sp(\gamma^5\gamma^\mu\gamma^\nu\gamma^\alpha\gamma^\beta) = 4i\epsilon^{\mu\nu\alpha\beta}$.

The dynamical information about the decay process are encrypted within the relativistic invariant form factors, so these parameters have been measured in non-perturbative approach.

Here, we consider the transition of an initial meson $q(m_2)\bar{q}(m_3)$ with mass $M_1$ to a final meson $q(m_1)\bar{q}(m_3)$ with mass $M_2$ induced by the quark transition $m_2 \to m_1$ through the current $\bar{q}(m_1)J_{\mu(\nu)}q(m_3)$. Now, in the decay mode $B \to (K, K^*)$, $m_1 = m_s$, $m_2 = m_b$, $m_3 = m_d$ and the structure of initial B meson vertex is like $\Gamma_1 = i\gamma_5 G_1/\sqrt{N_c}$, where $N_c = 1$ for leptons and $N_c = 3$ for quarks. For a pseudoscalar state, final meson vertex has the form $\Gamma_2 = i\gamma_5 G_2/\sqrt{N_c}$ and for an S-wave vector meson, the vertex is $\Gamma_{2\mu} = [A\gamma_\mu + B(k_1 - k_3)_\mu]G_2/\sqrt{N_c}$ with $A = -1$ and $B = 1/(\sqrt{s_2} + m_1 + m_3)$. The spectral representation of the form factors at $q^2 < 0$ could be written as [45],

$$f_i(q^2) = \frac{1}{16\pi^2} \int_{(m_1+m_3)^2}^{\infty} ds_2 \varphi_2(s_2) \int_{s_1^-(s_2,q^2)}^{s_1^+(s_2,q^2)} ds_1 \varphi_1(s_1) \frac{\tilde{f}_i(s_1, s_2, q^2)}{\lambda^{1/2}(s_1, s_2, q^2)}, \qquad (10)$$

where, the wave function $\varphi_i(s_i) = G_i(s_i)/(s_i - M_i^2)$ and

$$s_1^\pm(s_2, q^2) = \frac{s_2(m_1^2 + m_1^2 - q^2) + q^2(m_1^2 + m_3^2) - (m_1^2 - m_2^2)(m_1^2 - m_3^2)}{2m_1^2} \pm \frac{\lambda^{\frac{1}{2}}(s_2, m_3^2, m_1^2)\lambda^{\frac{1}{2}}(q^2, m_1^2, m_2^2)}{2m_1^2}. \qquad (11)$$

The triangle function is defined as, $\lambda(s_1, s_2, s_3) = (s_1 + s_2 - s_3)^2 - 4s_1 s_2$ and the function $\varphi(s)$ has the form

$$\varphi(s) = \frac{\pi}{\sqrt{2}} \frac{\sqrt{s^2 - (m_q^2 - m_{\bar{q}}^2)^2}}{\sqrt{s - (m_q - m_{\bar{q}})^2}} \frac{\omega(k^2)}{s^{3/4}}, \qquad (12)$$

where, $k = \lambda^{1/2}(s, m_q^2, m_{\bar{q}}^2)/2\sqrt{s}$ and $\omega(k^2)$ is the ground-state S-wave radial wave function and normalized as $\int_0^\infty dk k^2 |\omega(k^2)|^2 = 1$. The un-subtracted double spectral densities $\tilde{f}_i(s_1, s_2, q^2)$ could be found in ref. [60].

In dispersion quark model, only two particle $q\bar{q}$ intermediated states are taken into account to construct the amplitudes for the interaction between $q\bar{q}$ quark pair and external field. In this context, it is very much close to light-cone quark model (LCQM) [61]. The form factors of LCQM could be written as un-subtracted spectral densities or in the form of double spectral representation, which are analogous to the dispersion formulation at $q^2 < 0$. But, the LCQM form factors $f, a_1, a_2$ and $h$ are different from dispersion model in the next-to-leading $1/m_Q$ order.

To obtain the form factors which are more reliable at large $q^2$ and also compatible with the lattice results, quark masses and wave functions of Godfrey-Isgur (GI) [62] model with a switched off one-gluon exchange (OGE) are taken into consideration. This approach only contains the impact of the confinement scale. Table 3 represents the fitted values of form factors calculated from GI-OGE model.

**Table 3:** Parameters for GI-OGE model fit [63]

| Decay | $B \to K$ | | | $B \to K^*$ | | | | | | |
|---|---|---|---|---|---|---|---|---|---|---|
| | $f_+(0)$ | $f_-(0)$ | $s(0)$ | $g(0)$ | $f(0)$ | $a_+(0)$ | $a_-(0)$ | $h(0)$ | $g_+(0)$ | $g_-(0)$ |
| Para- | $\sigma_1$ | $\sigma_1$ | $\sigma_1$ | $\sigma_1$ | $\sigma_1$ | $\sigma_1$ | $\sigma_1$ | $\sigma_1$ | $\sigma_1$ | $\sigma_1$ |
| meter | $\sigma_2$ | $\sigma_2$ | $\sigma_2$ | $\sigma_2$ | $\sigma_2$ | $\sigma_2$ | $\sigma_2$ | $\sigma_2$ | $\sigma_2$ | $\sigma_2$ |
| GI-OGE | 0.33 | -0.27 | 0.057 | 0.063 | 2.01 | -0.0454 | 0.053 | 0.0056 | -0.354 | 0.313 |
| | 0.0519 | 0.0524 | 0.0517 | 0.0523 | 0.0212 | 0.039 | 0.044 | 0.0657 | 0.0523 | 0.053 |
| | 0.00065 | 0.00066 | 0.00064 | 0.00066 | 0.00009 | 0.00004 | 0.00023 | 0.0010 | 0.0007 | 0.00067 |

### ii) Expressions for different observables

In this section, we present formulas for differential decay rates, forward-backward asymmetries and lepton polarization asymmetries. The expressions for differential decay rates and forward-backward asymmetry agree to the formulation of ref [64] and also reproduce the formulas of ref [65, 66] for $m_s = 0$ and $m_l \neq 0$, while for $m_s = 0$ and $m_l = 0$ reproduce the formulas of ref [67]. In the limit $m_s = 0$, expressions for lepton polarization asymmetries corresponds to ref [63, 65, 66].

Introducing the dimensionless kinematical variables $\hat{s} = q^2/M_B^2$ and $\hat{t} = (P_B - p_{l+})^2/M_B^2$, double differential decay width for $B \to K l^+ l^-$ decay has the following form

$$\frac{d^2\Gamma(B \to K l^+ l^-)}{d\hat{s} d\hat{t}} = \frac{G_F^2 M_B^5 |V_{ts}^* V_{tb}|^2 \alpha_{em}^2}{265 \pi^5} \left[-\hat{\Pi}\beta_P + 2\hat{m}\delta_P\right] \tag{13}$$

where,

$$\beta_P = \left|C_{9V}^{eff}(m_b, q^2) f_+(q^2) + 2(m_b + m_s) C_{7\gamma}(m_b) s(q^2)\right|^2 + |C_{10A}(m_b) f_+(q^2)|^2$$

$$\hat{\Pi} = (\hat{t} - 1)(\hat{t} - \hat{r}) + \hat{s}\hat{t} + \hat{m}(1 + \hat{r} + \hat{m} - \hat{s} - 2\hat{t})$$

$$\delta_P = |C_{10A}|^2 \left\{\left(1 + \hat{r} - \frac{\hat{s}}{2}\right) |f_+(q^2)|^2 + (1 - \hat{r}) Re[f_+(q^2) f_-^*(q^2)] + \frac{\hat{s}}{2} |f_-(q^2)|^2\right\} \tag{14}$$

with, $\hat{r} \equiv (M_K/M_B)^2$ and $\hat{m} \equiv (m_l/M_B)^2$. Now, we integrate $\frac{d^2\Gamma}{d\hat{s}d\hat{t}}$ over $\hat{t}$ from $\hat{t}_{min} = \left[1 + \hat{r} + 2\hat{m} - \hat{s} - \sqrt{\left(1 - \frac{4\hat{m}}{\hat{s}}\right) \lambda^{1/2}(1, \hat{s}, \hat{r})}\right]/2$ to

$\hat{t}_{max} = \left[1 + \hat{r} + 2\hat{m} - \hat{s} + \sqrt{(1 - \frac{4\hat{m}}{\hat{s}}) \lambda^{1/2}(1, \hat{s}, \hat{r})}\right]/2$, where, $\lambda(1, \hat{s}, \hat{r}) = 1 + \hat{r}^2 + \hat{s}^2 - 2\hat{r} - 2\hat{s} - 2\hat{r}\hat{s}$, and we have got expression for dilepton mass distribution (or differential decay rate)

$$\frac{d\Gamma(B \to K l^+ l^-)}{d\hat{s}} = \frac{G_F^2 M_B^5 |V_{ts}^* V_{tb}|^2 \alpha_{em}^2}{1536\pi^5}\left[\sqrt{1-\frac{4\hat{m}}{\hat{s}}}\lambda^{\frac{1}{2}}(1,\hat{s},\hat{r})\beta_P + 12\hat{m}\delta_P\right]. \tag{15}$$

For the decay channel B → K*l+l− double differential decay width is given by

$$\frac{d^2\Gamma(B \to K^* l^+ l^-)}{d\hat{s}d\hat{t}} = \frac{G_F^2 M_B^5 |V_{ts}^* V_{tb}|^2 \alpha_{em}^2}{512\pi^5}\left[\beta_V^{(1)} + \beta_V^{(2)} + 4\hat{m}\delta_V\right] \tag{16}$$

where,

$$\beta_V^{(1)} = [(\hat{s}+2\hat{m})\lambda(1,\hat{s},\hat{r}) + 2\hat{s}\hat{\Pi}]|G(q^2)|^2 + \left[\hat{s}+2\hat{m}-\frac{\hat{\Pi}}{2\hat{r}}\right]|F(q^2)|^2$$
$$- \frac{\lambda^2(1,\hat{s},\hat{r})}{2\hat{r}}\hat{\Pi}|H_+(q^2)|^2 + \frac{\hat{s}-1+\hat{r}}{\hat{r}}\hat{\Pi}R(q^2)$$

$$\beta_V^{(2)} = 2\hat{s}[2\hat{t}+\hat{s}-\hat{r}-1-2\hat{m}]R_1(q^2)$$

$$|G(q^2)|^2 = \left|C_{9V}^{eff}(m_b,q^2)M_B g(q^2) - \frac{2C_{7\gamma}(m_b)}{\hat{s}}\frac{(m_b+m_s)}{M_B}g_+(q^2)\right|^2 + |C_{10A}(m_b)M_B g(q^2)|^2$$

$$|F(q^2)|^2 = \left|C_{9V}^{eff}(m_b,q^2)\frac{f(q^2)}{M_B} - \frac{2C_{7\gamma}(m_b)}{\hat{s}}\frac{(m_b-m_s)}{M_B}(1-\hat{r})B_0(q^2)\right|^2 + \left|C_{10A}(m_b)\frac{f(q^2)}{M_B}\right|^2$$

$$|H_+(q^2)|^2 =$$
$$\left|C_{9V}^{eff}(m_b,q^2)M_B a_+(q^2) - \frac{2C_{7\gamma}(m_b)}{\hat{s}}\frac{(m_b-m_s)}{M_B}B_+(q^2)\right|^2 + |C_{10A}(m_b)M_B a_+(q^2)|^2$$

$$R(q^2) =$$
$$Re\left\{\left[C_{9V}^{eff}(m_b,q^2)\frac{f(q^2)}{M_B} - \frac{2C_{7\gamma}(m_b)}{\hat{s}}\frac{(m_b-m_s)}{M_B}(1-\hat{r})B_0(q^2)\right]\left[C_{9V}^{eff}(m_b,q^2)M_B a_+(q^2) - \frac{2C_{7\gamma}(m_b)}{\hat{s}}\frac{(m_b-m_s)}{M_B}B_+(q^2)\right]^*\right\} + |C_{10A}(m_b)|^2 Re[a_+(q^2)f^*(q^2)]$$

$$R_1(q^2) = Re\left\{\left[C_{9V}^{eff}(m_b,q^2)M_B g(q^2) - \frac{2C_{7\gamma}(m_b)}{\hat{s}}\frac{(m_b+m_s)}{M_B}g_+(q^2)\right]\left[C_{10A}(m_b)\frac{f(q^2)}{M_B}\right]^*\right\} +$$
$$Re\left\{\left[C_{9V}^{eff}(m_b,q^2)\frac{f(q^2)}{M_B} - \frac{2C_{7\gamma}(m_b)}{\hat{s}}\frac{(m_b-m_s)}{M_B}(1-\hat{r})B_0(q^2)\right][C_{10A}(m_b)M_B g(q^2)]^*\right\}$$

$$B_0(q^2) = g_+(q^2) + g_-(q^2)\frac{\hat{s}}{1-\hat{r}}$$

$$B_+(q^2) = -\hat{s}M_B^2 \frac{h(q^2)}{2} - g_+(q^2)$$

$$\delta_V = \frac{|C_{10A}|^2}{2}\lambda(1,\hat{s},\hat{r})\{-2|g(q^2)M_B|^2 - \frac{3}{\lambda(1,\hat{s},\hat{r})}\left|\frac{f(q^2)}{M_B}\right|^2 + \frac{2(1+k)-\hat{s}}{4\hat{r}}|a_+(q^2)M_B|^2 +$$
$$\frac{\hat{s}}{4\hat{r}}|a_-(q^2)M_B|^2 + \frac{1}{2\hat{r}}Re[f(q^2)a_+^*(q^2) + f(q^2)a_-^*(q^2)] + \frac{1-\hat{r}}{2\hat{r}}Re[M_B a_+(q^2)M_B a_-^*(q^2)]\}$$

$$\beta_V = 2\lambda(1,\hat{s},\hat{r})\hat{s}|G(q^2)|^2 + \left[2\hat{s}+\frac{(1-\hat{r}-\hat{s})^2}{4\hat{r}}\right]|F(q^2)|^2 + \frac{\lambda^2(1,\hat{s},\hat{r})}{4\hat{r}}|H(q^2)|^2$$
$$+\frac{\lambda(1,\hat{s},\hat{r})}{2\hat{r}}(\hat{s}-1+\hat{r})R(q^2) \tag{17}$$

Here, $\hat{r} \equiv (M_{K^*}/M_B)^2$. After integrating over the variable $\hat{t}$, the differential decay rate can be written in the form

$$\frac{d\Gamma(B \to K^* l^+ l^-)}{d\hat{s}} = \frac{G_F^2 M_B^5 |V_{ts}^* V_{tb}|^2 \alpha_{em}^2}{1536\pi^5} \sqrt{1 - \frac{4\hat{m}}{\hat{s}}} \lambda^{\frac{1}{2}}(1, \hat{s}, \hat{r}) [\beta_V (1 + \frac{2\hat{m}}{\hat{s}}) + 12\hat{m}\delta_V]. \quad (18)$$

The expression for the differential decay rates for $B \to (K, K^*) \nu \bar{\nu}$ can be obtained from the corresponding formulas of $B \to (K, K^*) l^+ l^-$ by the following replacements

$$\hat{m} \to 0, C_{7\gamma} \to 0, C_{9V}^{eff} \to \frac{X(x_t)}{\sin^2(\theta_W)}, C_{10A} \to -\frac{X(x_t)}{\sin^2(\theta_W)}. \quad (19)$$

For $\to K^* l^+ l^-$, the forward-backward (FB) asymmetry is defined as,

$$A_{FB}(\hat{s}) = \frac{1}{d\Gamma(B \to K^* l^+ l^-)/d\hat{s}} \left[ \int_0^1 d\cos\theta \frac{d^2\Gamma(B \to K^* l^+ l^-)}{d\hat{s}\, d\cos\theta} - \int_{-1}^0 d\cos\theta \frac{d^2\Gamma(B \to K^* l^+ l^-)}{d\hat{s}\, d\cos\theta} \right], \quad (20)$$

where, $\theta$ is the angle between the final state lepton $l^+$ and the B-meson in dilepton rest frame. FB asymmetry is quite interesting property for any decay channel as it is sensitive to parity status of any interaction. At low $q^2$ region, parity conserving photonic interaction is relatively dominant leading to a small FB asymmetry. But in higher momentum region (i.e. large $q^2$), parity violating Z- and W-boson contributions become more significant. As a consequence FB asymmetry becomes larger. This phenomenon has also a relevant connection with the flipping of sign of Wilson coefficients. The explicit expression for FB asymmetry is given as

$$A_{FB} = \frac{3\hat{s}\sqrt{1 - 4\hat{m}/\hat{s}} \lambda^{1/2}(1, \hat{s}, \hat{r}) R_1(q^2)}{(1 + 2\hat{m}/\hat{s})\beta_V + 12\hat{m}\delta_V}. \quad (21)$$

Within the SM, $A_{FB}$ for $B \to K l^+ l^-$ is very small [8] because the hadronic current for $B \to K$ transition does not have any axial-vector contribution. It can have a non-zero value only in the presence of NP. But NP in the form of vector/ axial-vector operators cannot enhance the measurement whereas NP in the form of scalar/ pseudo-scalar or tensor operators can increase the percent level of $A_{FB}$. If the increment is only in high $q^2$ region, scalar/ pseudo-scalar NP operators does not play any key role. But, when it shows a significant enhancement at low $q^2$ and gradually rises with increasing $q^2$, scalar/ pseudo-scalar NP operators are highly presumable. The intermediate region i.e. $7 < q^2 < 12$ GeV$^2$ is highly sensitive to the tensor NP operators [68]. In our case, the model does not contain any scalar/ pseudo-scalar or tensor operators, so $A_{FB}$ does not undergo any NP contribution for $B \to K l^+ l^-$ channel.

In order to find out the longitudinal lepton polarization asymmetry, we can define it in the following way,

$$P_L(\hat{s}) = \frac{1}{d\Gamma/d\hat{s}} \left[ \frac{d\Gamma(h = -1)}{d\hat{s}} - \frac{d\Gamma(h = +1)}{d\hat{s}} \right], \quad (22)$$

where, $h = -1(+1)$ describes right- (left-) handed charged lepton in the final state.

For $B \to K l^+ l^-$ channel, we have

$$P_L(\hat{s}) = \frac{2\sqrt{1 - \frac{4\hat{m}}{\hat{s}}} \lambda(1, \hat{s}, \hat{r})}{\left(1 + \frac{2\hat{m}}{\hat{s}}\right) \lambda(1, \hat{s}, \hat{r}) \beta_P + 12\hat{m}\delta_P}$$
$$Re\{[C_{9V}^{eff}(m_b, q^2) f_+(q^2) + 2(m_b + m_s) C_{7\gamma}(m_b) s(q^2)] C_{10A}^* f_+^*(q^2)\} \tag{23}$$

and for the process $B \to K^* l^+ l^-$, $P_L$ is given as

$$P_L(\hat{s}) = \frac{2\sqrt{1 - \frac{4\hat{m}}{\hat{s}}}}{\left(1 + \frac{2\hat{m}}{\hat{s}}\right) \beta_V + 12\hat{m}\delta_V} [2\lambda(1, \hat{s}, \hat{r}) \hat{s} R_G(q^2) + \left(2\hat{s} + \frac{(1 - \hat{r} - \hat{s})^2}{4\hat{r}}\right) R_F(q^2)$$
$$+ \frac{\lambda^2(1, \hat{s}, \hat{r})}{4\hat{r}} R_{H_+}(q^2) - \frac{\lambda(1, \hat{s}, \hat{r})}{4\hat{r}} (\hat{s} - 1 + \hat{r}) R_R(q^2)] \tag{24}$$

$$R_G(q^2) = Re\left\{[C_{9V}^{eff}(m_b, q^2) M_B g(q^2) - \frac{2C_{7\gamma}(m_b)}{\hat{s}} \frac{(m_b + m_s)}{M_B} g_+(q^2)][C_{10A}(m_b) M_B g(q^2)]^*\right\}$$

$$R_F(q^2) = Re\left\{[C_{9V}^{eff}(m_b, q^2) \frac{f(q^2)}{M_B} - \frac{2C_{7\gamma}(m_b)}{\hat{s}} \frac{(m_b - m_s)}{M_B} (1 - \hat{r}) B_0(q^2)] [C_{10A}(m_b) \frac{f(q^2)}{M_B}]^*\right\}$$

$$R_{H_+}(q^2) =$$
$$Re\left\{[C_{9V}^{eff}(m_b, q^2) M_B a_+(q^2) - \frac{2C_{7\gamma}(m_b)}{\hat{s}} \frac{(m_b - m_s)}{M_B} B_+(q^2)] [C_{10A}(m_b) M_B a_+(q^2)]^*\right\}$$

$$R_R(q^2) =$$
$$Re\left\{[C_{9V}^{eff}(m_b, q^2) \frac{f(q^2)}{M_B} - \frac{2C_{7\gamma}(m_b)}{\hat{s}} \frac{(m_b - m_s)}{M_B} (1 - \hat{r}) B_0(q^2)] [C_{10A}(m_b) M_B a_+(q^2)]^*\right\}$$
$$+ Re\left\{[C_{9V}^{eff}(m_b, q^2) M_B a_+(q^2) - \frac{2C_{7\gamma}(m_b)}{\hat{s}} \frac{(m_b - m_s)}{M_B} B_+(q^2)] [C_{10A}(m_b) \frac{f(q^2)}{M_B}]^*\right\} \tag{25}$$

## 4 Results and Discussion

### i) Contribution of $Z'$ to various observables

The impact of non-universal $Z'$ model to different observables of $B \to (K, K^*) l^+ l^- (\nu \bar{\nu})$ has been discussed in this section. As $b \to s e^+ e^-$ transitions are much constricted from various experimental bounds, it is desirable that the NP effect comes from only $b \to s \mu^+ \mu^-$ and $b \to s \tau^+ \tau^-$ channels. So, to have fruitful results we have implicated our model only to $\mu$- and $\tau$-channels by replacing $C_9^{eff} \to C_9^{Total}$ and $C_{10} \to C_{10}^{Total}$.

### a. Branching Ratio

We have predicted the branching ratio for $B \to (K, K^*) l^+ l^- (\nu \bar{\nu})$ where $(l = \mu, \tau)$ and summarized our results in Table 4. The distributions for differential branching ratio in Fig. 1 indicate some appealing phenomena of the corresponding channels. As we see, resonant contributions coming from $c\bar{c}$ loop are highly peaked in narrow regions around their masses. So, outside of these regions one can neglect their effects. For $\mu$-channels both the peaks for $J/\psi$ and $\psi'$ are present but in $\tau$-channels we observe only $\psi'$ peak within the kinematical region. Here we notice one more interesting fact that the amplitude for $B \to K^* l^+ l^-$ has a kinematical pole at $q^2 = 0$ whereas the amplitude for $B \to K l^+ l^-$ is regular at $q^2 = 0$. This makes $B \to K^*$ transitions very sensitive at low $q^2$ region. Thus, to predict accurate results for branching ratio one needs to gather precise knowledge about form factors at low $q^2$. On the other hand $B \to K$ transitions are more stable with respect to variation of relevant form factors. In our calculation we have combined the values of branching ratios for low $q^2$ and high $q^2$ region and excluded the intermediate regions i.e. $7.4 < q^2 < 14.6$ GeV$^2$ for $B \to K\mu^+\mu^-$ and $9.5 < q^2 < 13.6$ GeV$^2$ for $B \to K^*\mu^+\mu^-$ channels. For $\tau$-channels the prediction lies in the range $14 < q^2 < q^2_{max}$.

For each of the modes we have seen noticeable effect of $Z'$. We have computed branching ratios for the three scenarios as we discussed in section **2.ii)** and summarized the result in Table 4.

**Table 4:** Branching ratios of $B \to (K, K^*) l^+ l^- (\nu \bar{\nu})$ decay channels

| | Decay modes | SM prediction | Expt. Result [69] | $Z'$ model | | |
|---|---|---|---|---|---|---|
| | | | | $S_1$ | $S_2$ | $S_3$ |
| $\mathcal{BR} \times 10^{-7}$ | $B \to K\mu^+\mu^-$ | 3.50±0.6 | 3.39±0.34 | 4.83±0.29 | 4.15±0.36 | 1.69±0.94 |
| | $B \to K\tau^+\tau^-$ | 1.01±0.06 | - | 1.59±0.04 | 1.27±0.08 | 0.42±0.41 |
| $\mathcal{BR} \times 10^{-6}$ | $B \to K^*\mu^+\mu^-$ | 1.16±0.61 | 0.94±0.05 | 1.42±0.08 | 1.16±0.15 | 0.49±0.29 |
| | $B \to K^*\tau^+\tau^-$ | 0.18±0.01 | - | 0.26±0.02 | 0.21±0.03 | 0.008±0.006 |
| | $B \to K\nu\bar{\nu}$ | 2.43±0.09 | < 4.9×10$^{-5}$ | 2.98±0.57 | 2.77±0.46 | 2.69±0.37 |
| | $B \to K^*\nu\bar{\nu}$ | 6.57±0.36 | < 5.5×10$^{-5}$ | 8.05±1.82 | 7.49±1.14 | 7.29±0.98 |

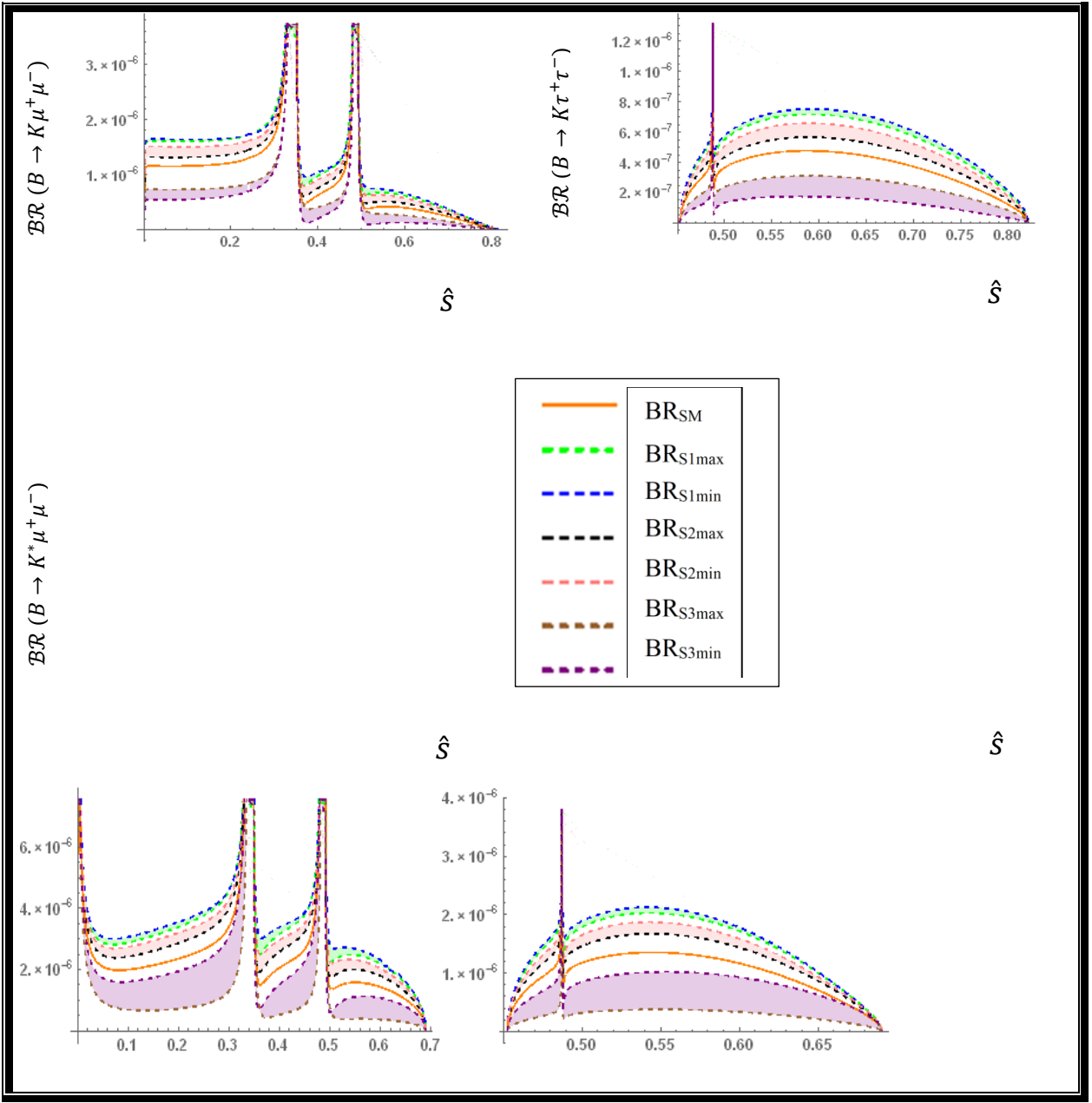

**Fig. 1: (a)** Differential Branching ratio for $B \to (K, K^*) l^+ l^-$ decay modes

Those channels which include neutrinos (e.g. $B \to (K, K^*)\nu\bar{\nu}$) as final state particles instead of lepton pairs are out of all such contaminated facts. For this reason we have computed our results for full kinematical region i.e. $q^2_{min} < q^2 < q^2_{max}$. The Wilson coefficients are replaced according to the relations written in eq. (8).

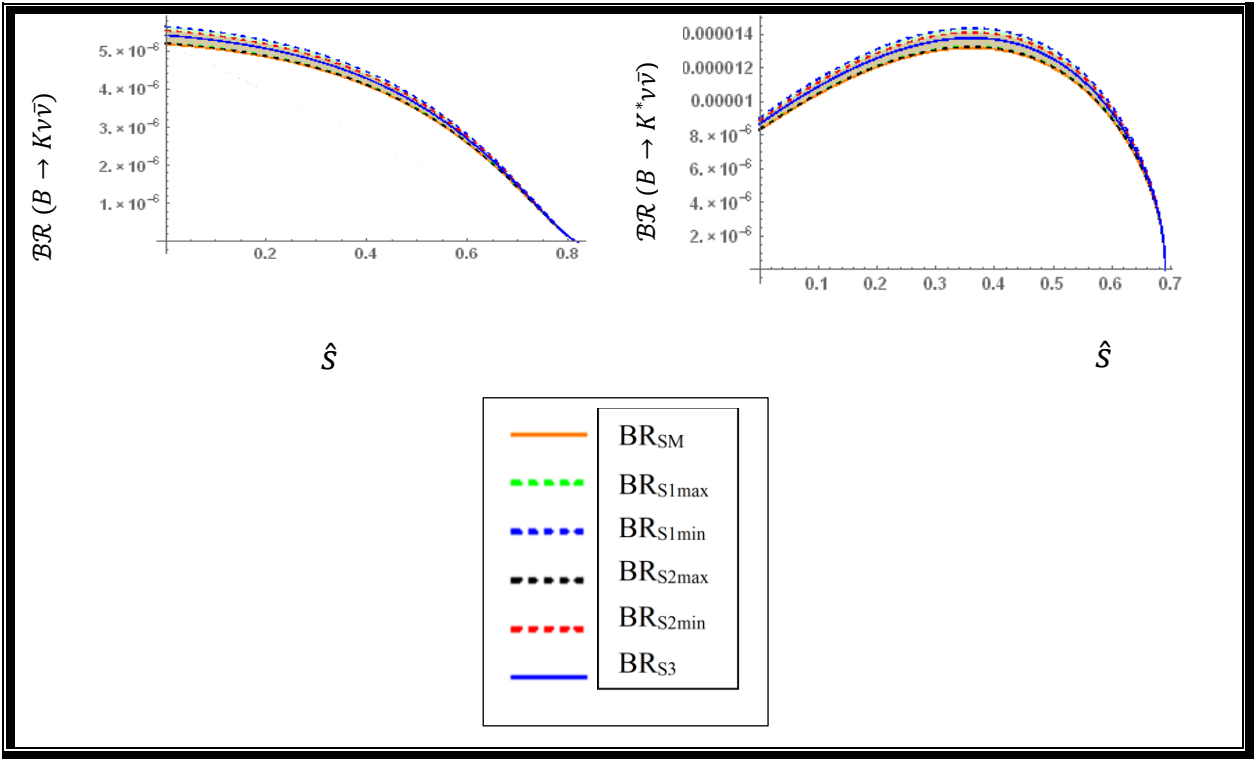

**Fig. 1: (b)** Differential Branching ratio for $B \to (K, K^*)\nu\bar{\nu}$ decay modes

Corresponding to the above table and figures, it is clear that, all the decay channels described in this paper get the maximum branching fraction for lowest value of $B_{sb}$ and the highest values of $S_{LL}$ and $D_{LL}$ (sign of both $S_{LL}$ and $D_{LL}$ is negative in this case ). So one can get higher probability of these $B \to K$ and $B \to K^*$ transitions by considering small quarks coupling and large lepton couplings with $Z'$ boson.

We have observed another important fact from our prediction that the bands for third scenario of each case (except neutrino channels) go down from SM curve whereas other two scenarios make the branching ratio higher than the SM. The reason behind such a fascinating behaviour might come from the coupling of quarks with $Z'$ as well as from the weak phase angle because $S_{LL}$ and $D_{LL}$ have constant values in this scenario. Here we found out that as we increase the quark coupling $B_{sb}$ and decrease the phase angle $\phi_{sb}$ (has negative sign), branching fraction for $B \to K^* l^+ l^-$ channels reaches to least value whereas for $B \to K l^+ l^-$ channels, it is lowest for smaller $B_{sb}$ and higher $\phi_{sb}$.

### b. Forward-Backward Asymmetry

The forward-backward asymmetries for $B \to K^* l^+ l^-$ ($l = \mu, \tau$) channels are shown in Fig 2. We have observed a strong sensitivity of the asymmetries with relevant form factors as well as the sign of Wilson coefficients. From the figures we see that in low $\hat{s}$ region the shape of asymmetry is very much influenced by the inversion of the signs of Wilson coefficients compared to the SM. For $\mu$-channels, when all coefficients are used according to SM values, we observe that maximum asymmetry is found within $q^2/M_B^2 \simeq 0.1$ above which one can find the zero crossing. If we use the conversion rule $C_{7\gamma}^{SM} \to -C_{7\gamma}$, $A_{FB}$ remains negative for whole regime of $\hat{s}$ whereas for $C_{9V}^{SM} \to -C_{9V}$, $A_{FB}$ is fully positive excluding the uncertainty regions. These two conversions show no zero crossing below the $c\bar{c}$ resonances and the latter case is much satisfying with experimental results. For $\tau$-channels, the shape of $A_{FB}$ is same for $C_{7\gamma}^{SM}$ and $-C_{7\gamma}$, for both the cases $A_{FB}$ remains negative while for $C_{9V}^{SM} = -C_{9V}$, the asymmetry shifts to positive quadrant retaining the maximum absolute value ~0.15 same as SM predictions. The effect of NP i.e. $Z'$ contribution to each of the above cases are well consisting with SM estimations. Only the maximum values of parameters for third scenario show a noticeable variation from other prediction. For each condition taken above, maximum parametric values of third scenario flip the shape of asymmetry. This contradictory fact might come due to the coupling of quarks and leptons with $Z'$ boson. When the value of $B_{sb}$ is the highest and $S_{LL}$ is smaller than $D_{LL}$ (negative) the nature of forward-backward asymmetry changes its shape. So, to have a good prediction of $A_{FB}$ corresponding to the above sign flip rules, one should avoid larger coupling of quarks with $Z'$.

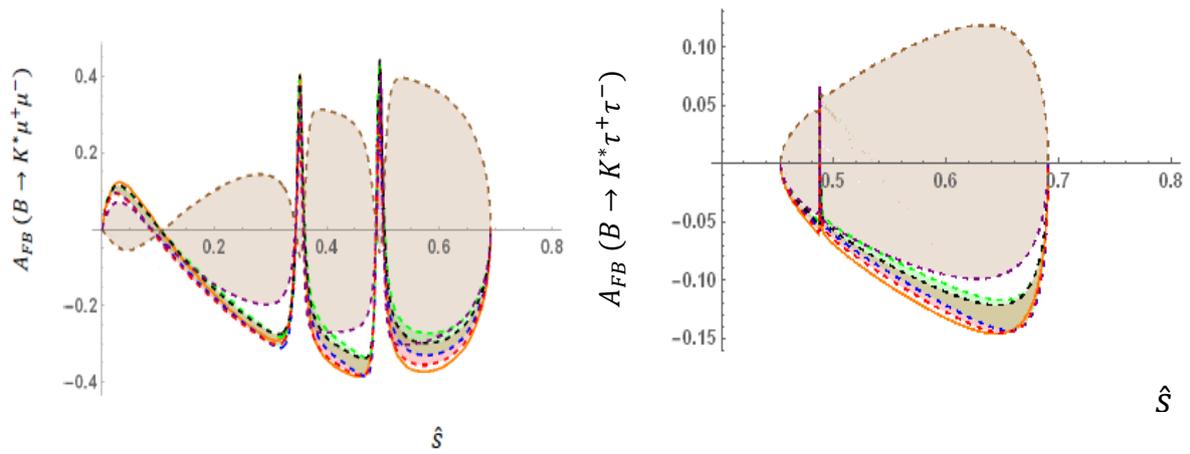

**a) (i)**  **a) (ii)**

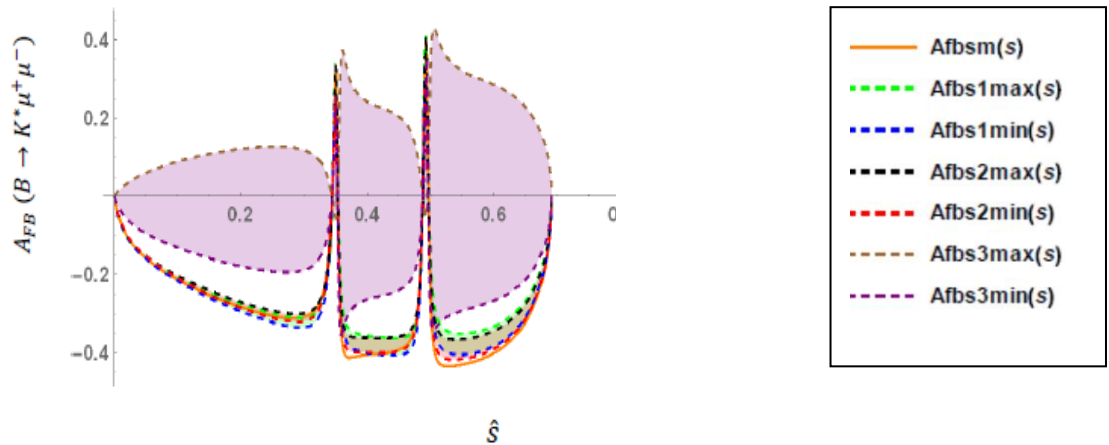

**b) (i)**

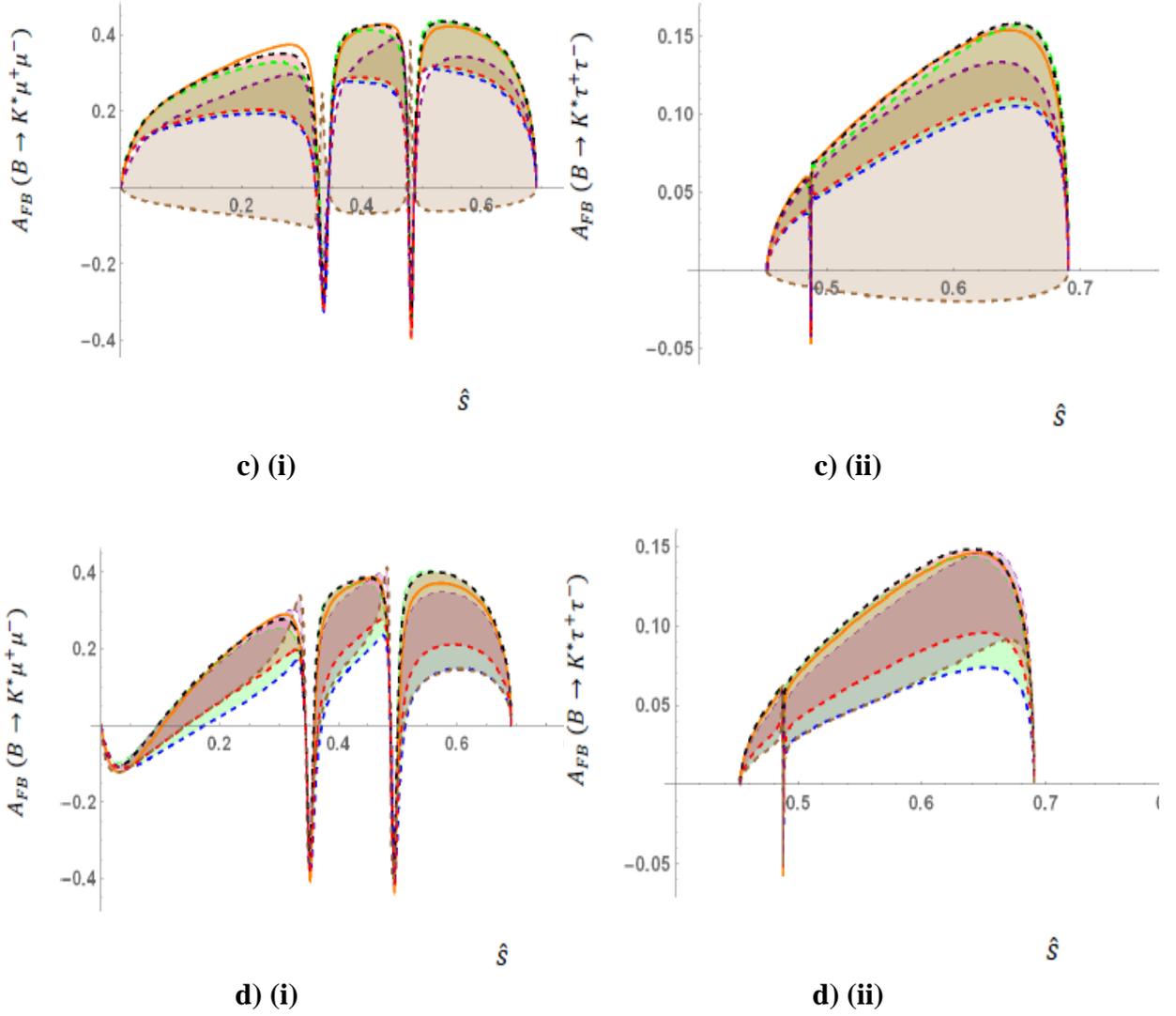

**Fig. 2:** Forward-Backward Asymmetry for $B \to K^* l^+ l^-$ ($l = \mu, \tau$) for different conditions: **a) (i, ii)** taking $C_7^{SM}, C_9^{SM}, C_{10}^{SM}$; **b) (i)** taking $C_7 \to -C_7^{SM}$; **c) (i, ii)** taking $C_9 \to -C_9^{SM}$; **d) (i, ii)** taking $C_{10} \to -C_{10}^{SM}$.

Besides these estimations we have adopted another sign flipping relation i.e. $C_{10}^{SM} = -C_{10}$ and found that for both $\mu$- and $\tau$-channels, the shape of $A_{FB}^{SM}$ is totally flipped for full kinematic region from other two conditions discussed before. The NP contribution in this case does not change the shape of asymmetry at all but drives zero point towards a higher value of $q^2/M_B^2$ than the SM. After compiling the $Z'$ contribution with the SM, we have found that the maximum value of $A_{FB}$ for $\mu$-channel is $\sim 0.4$ and for $\tau$-channel it is $\sim 0.15$.

### c. Lepton Polarization Asymmetry

To understand the behaviour of lepton polarization asymmetry $P_L$, we should look towards the following figures. For $B \to K l^+ l^-$ channels, $P_L$ vanishes at threshold due to the presence of the kinematical factor $\sqrt{1 - 4\hat{m}/\hat{s}}$. In the intermediate region of $q^2$, $P_L$ for $\mu$-channel goes down to the SM value $P_L = 2C_9 C_{10}/(C_9^2 + C_{10}^2) \simeq -1$ whereas for $B \to K\tau^+\tau^-$ mode $P_L$ does not go down to the value $\simeq -1$. Without any dependence on any particular $B \to K$ form factors or on long distance effects, $P_L$ is negative for whole kinematical accessible region of $q^2$ and only in the resonance area i.e. $J/\psi, \psi'$ mass region, $P_L$ becomes positive for $\mu$-channel (only $\psi'$ for $\tau$-channel). A weak $q^2$ dependence of the non-resonant $P_L$ is due to the perturbative part present in $C_9^{eff}$.

For $B \to K^* l^+ l^-$ channel, $P_L$ is not behaving in the exactly same manner. The term in $H_{eff}$ proportional to a small $C_{7\gamma}$ consists a photon pole at $q^2 = 0$ and thus a parity-conserving photon exchange dominates the decay at low $q^2$ region providing small value of $P_L$. As we can see from Fig.3, due to large enhancement of $P_L$ in all three scenarios, $Z'$ contribution to lepton polarization asymmetry is highly appreciated in our study. One intriguing fact coming out from the study of $P_L$ mounts our prediction to new level. We have witnessed that in third scenario, the polarization asymmetry gains positive nature for $B \to K^* \mu^+ \mu^-$ channel in the high recoil region before resonance loop effect. This is possible only when the quarks participating in that transition have stronger coupling to $Z'$ and $S_{LL}$ is smaller than $D_{LL}$ (negative). On the other side, the same condition makes $P_L \sim 0$ for $B \to K^* \tau^+ \tau^-$ channel.

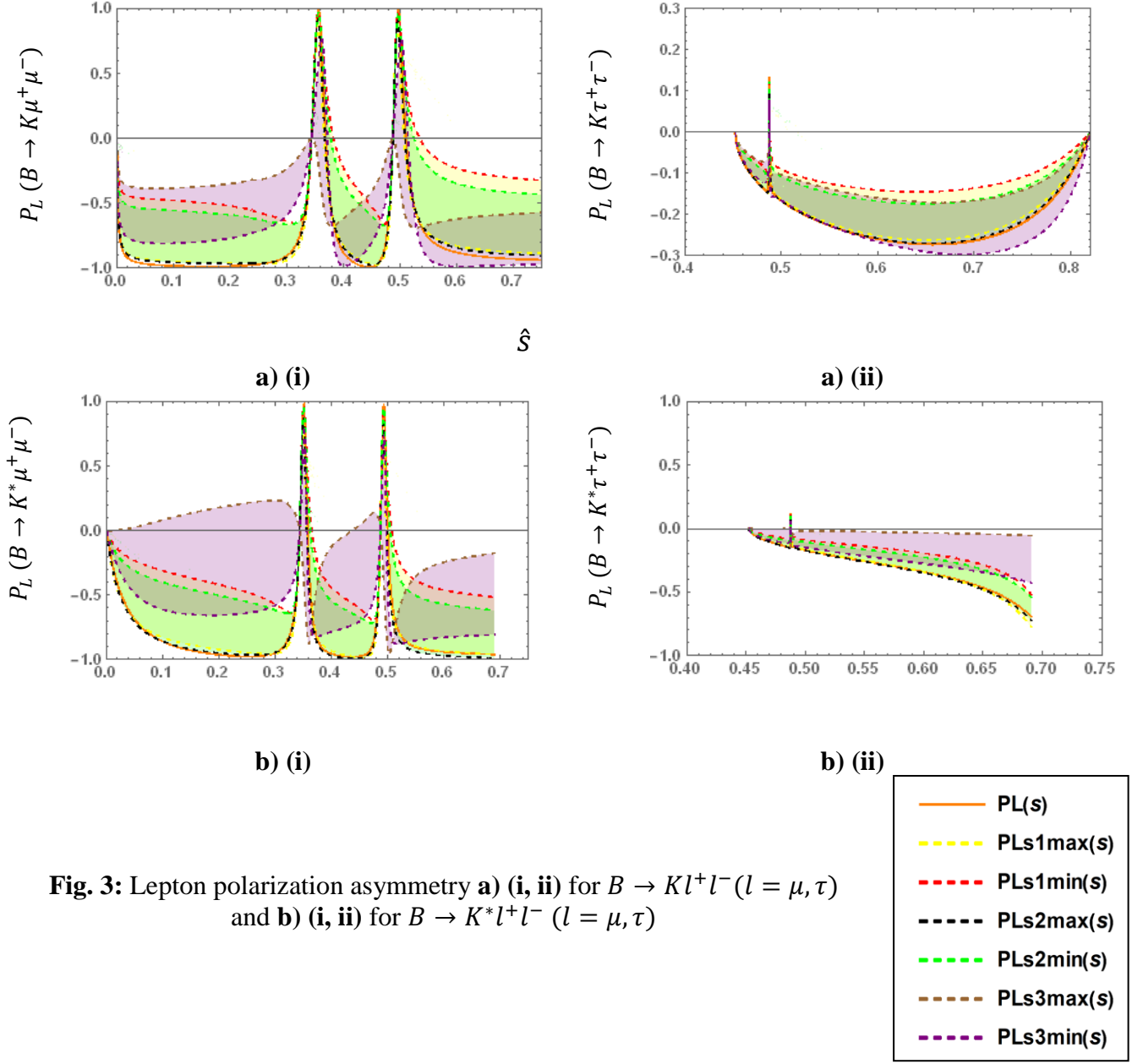

**Fig. 3:** Lepton polarization asymmetry **a) (i, ii)** for $B \to K l^+ l^- (l = \mu, \tau)$ and **b) (i, ii)** for $B \to K^* l^+ l^- (l = \mu, \tau)$

### ii) Explanation of $R_K$ and $R_{K^*}$

The best thing to consider non-universal $Z'$ model is that it is quite significant to the recent experimental evidences of $\mu - e$ universality violation observed in various colliders and accelerators. From the definition it is clear that $R_K$ and $R_{K^*}$ are ratios of branching fraction of $\mu$-channel to that of $e$-channel for $B \to K$ and $B \to K^*$ transitions respectively. According to the SM, all leptons are coupled to gauge bosons in similar way and the flavours always remain conserved. This fact leads us to the consequent that there is no non-universality between $\mu$ and $e$ and thus these ratios are unity within SM. But, experimentally observed values are quite different from SM expectation.

In our model, we have implemented the new chiral couplings of $Z'$ with the $\mu$-leptons only as $b \to s e^+ e^-$ are much constrained from experimental point of view. In our prediction we have found that to reach the experimentally measured values of both $R_K$ and $R_{K^*}$ the third scenario of $Z'$ model is more helpful than other two cases. The calculations have been done with different consideration of third scenario and the results are captured in Table 5.

**Table 5:** Values of $R_K$ and $R_{K^*}$ for third scenario ($S_3$) of Table 2

| Parameter | Bin size | $S_{LL}, D_{LL} \neq 0$ | $S_{LL} = 0$ | $D_{LL} = 0$ | Expt. Results |
|---|---|---|---|---|---|
| $R_K$ | 1 - 6 GeV$^2$ | 0.468±0.231 | 0.702±0.248 | 0.778±0.061 | $0.745^{+0.090}_{-0.074}$ |
| $R_{K^*}$ | 0.045-1.1 GeV$^2$ | 0.75±0.11 | 0.81±0.15 | 0.89±0.01 | $0.66^{+0.11}_{-0.07}$ |
|  | 1.1-6 Gev$^2$ | 0.50±0.31 | 0.71±0.32 | 0.89±0.05 | $0.69^{+0.11}_{-0.07}$ |

The results above depict a useful fact about the dependency of $R_K$ and $R_{K^*}$ to the Wilson coefficients. For $R_K$, when both $C_9^{Z'}$ and $C_{10}^{Z'}$ are present, results are smaller than experimental result. But when we put either $S_{LL} = 0$ or $D_{LL} = 0$, results are satisfying with the experimental result. So, we can say that NP effect is better noticeable when only one Wilson coefficient is modified among the two.

To get significant values of $R_{K^*}$ we have taken the same considerations as $R_K$ and found that at low $q^2$, both the coefficients $C_9^{Z'}$ and $C_{10}^{Z'}$ help combiningly to reach the experimental mean value of $R_{K^*}$. In central $q^2$ region, $S_{LL} = 0$ gives us better results than $S_{LL}, D_{LL} \neq 0$. On the other hand $D_{LL} = 0$ moves $R_{K^*}$ outside from the experimentally obtained region.

We have shown the correlation between $R_K$ and $R_{K^*}$ over the kinematical range of $1 < q^2 < 22$ GeV$^2$ in Fig.4 (a, b, c). Each point observed in the $R_K - R_{K^*}$ plane detects the value of these two lepton non-universality factors. There are three plots, all of which include the parametric data taken from the third scenario of $Z'$ model but with different conditions same as the Table 5. The horizontal and vertical bands detect the experimental range and the blue dashed line within the bands present the mean experimental value of the corresponding factors (data taken from the last two column of Table 5). Black solid lines are the SM values of $R_K$ and $R_{K^*}$. As the experimental values are smaller than the SM values, so we can restrict our observed points within lower left corner of the graphs. From the plotting below it is clear that only the minimum condition of scenario three holds within the restriction.

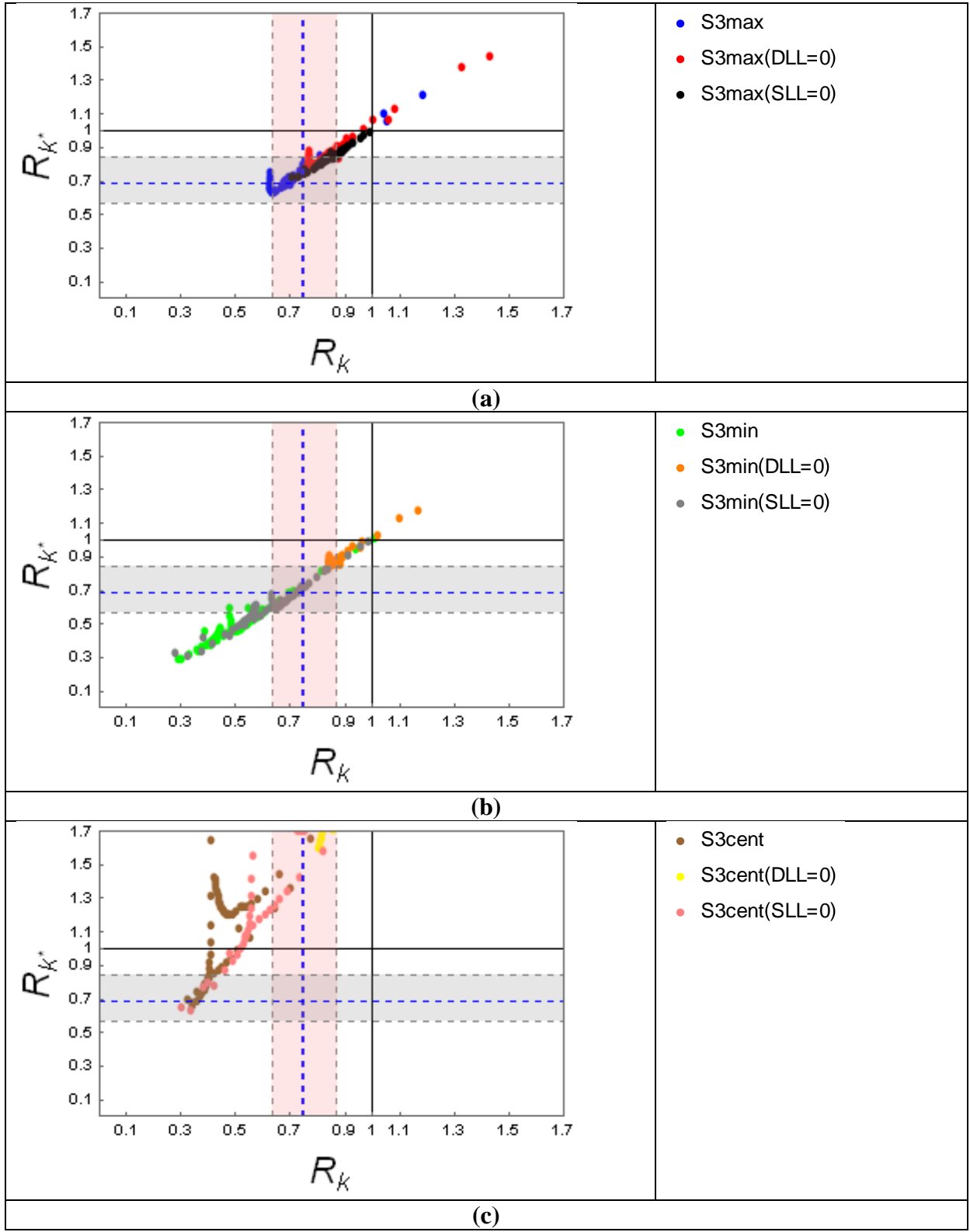

**Fig. 4:** Correlation between $R_K$ and $R_{K^*}$ **(a)** for maximum parametric values of $\mathcal{S}_3$, **(b)** for minimum parametric values of $\mathcal{S}_3$ and **(c)** for central parametric values of $\mathcal{S}_3$.

# 5  Conclusion

In this paper, we have presented our analysis of $B \to (K, K^*)(l^+l^-, \nu\bar{\nu})$ channels considering NP effect from non-universal $Z'$ model. We have studied the form factor dependent (FFD) observables such as $\frac{d\Gamma}{dq^2}$, $A_{FB}$, $P_L$ and lepton non-universality factors $R_K$ and $R_{K^*}$ incorporating the NP model. Our study produces several new motivations to search for the theory beyond SM and our predicted results are in good agreement with the experimental measurements. We have analysed some fascinating behaviour of the observables with the quark and lepton couplings and also with the weak phase angle. We have observed that in the presence of $Z'$ boson, all the observables have some differences from their SM results. Our main conclusions are summarized as:

- From fig. 1 (a) it could be seen that, both the processes $B \to K l^+l^-$ and $B \to K^* l^+l^-$ are sensitive to the NP couplings. From the current measurements of $B \to (K, K^*)\mu^+\mu^-$ modes it is found that the experimental values of branching ratio are less than their SM values. The non-universal $Z'$ model is in agreement with this fact for one of the three scenarios as the predictions depend on the interaction strength or couplings of $Z'$ with leptons as well as quarks. In future, our predictions could be more precise after the confirmation of $Z'$ boson.
- The prediction for $B \to (K, K^*) \nu\bar{\nu}$ branching fractions in this paper are also found to be satisfied with the experimental upper limits. The right handed coupling for neutrino is zero, so we get the maximum predicted branching ratio when left handed lepton coupling($B_{ll}^L \simeq 8 \times 10^{-2}$) is much higher.
- From our study of $A_{FB}$, we can conclude that the shape of this asymmetry is very useful to measure the signs of Wilson coefficients $C_{7\gamma}, C_{9V}, C_{10A}$. In the non-universal $Z'$ model, a larger $|B_{sb}|$ and a smaller $|\phi_{sb}|$ (negative) with $C_{9V} = -C_{9V}^{SM}$ are useful to account the discrepancy of experimentally measured $A_{FB}$ with that of the SM. This model predicts a shift of zero point towards higher $q^2$ with the sign flip of $C_{10A} = -C_{10A}^{SM}$.
- The longitudinal lepton polarization asymmetry for both $B \to (K, K^*) l^+l^-$ transitions does not depend on long-distance contribution at large $q^2$. $Z'$ contribution enhances the value of $P_L$ which is $\simeq -1$ within the SM. Positive nature of $P_L$ could also be seen from our prediction for the third case among the considerations which we have taken in our NP analysis.
- In our non-universal $Z'$ model, we are able to satisfy the experimental values of $R_K$ and $R_{K^*}$ in some particular considerations. It is expected that the currently observed anomalies in the $R_K$ and $R_{K^*}$ will be verified with the increased sensitivity of the LHCb detector and the Belle II experiment in the near future. The origin of the anomalies is expected to be discovered within the next few years which will enlighten us about the NP beyond the SM.

# Acknowledgment

P. Maji acknowledges the DST, Govt. of India for providing INSPIRE Fellowship through IF160115 for her research work. S. Sahoo and P. Nayek would like to thank SERB, DST, Govt. of India for financial support through grant no. EMR/2015/000817. S. Sahoo also acknowledges the financial support of NIT Durgapur through "Research Initiation Grant" office order No. NITD/Regis/OR/25 dated 31$^{st}$ March, 2014 & NITD/Regis/OR/2014 Dated 12$^{th}$ August, 2014.